\documentclass[12pt,letterpaper]{article}

\usepackage{iftex}
\ifPDFTeX
  \usepackage[T1]{fontenc}
  \usepackage{newtxtext,newtxmath}
\else
  \usepackage{fontspec}
  \usepackage{newtxmath}
\fi

\usepackage[
  top=1in,
  bottom=1in,
  left=0.75in,
  right=0.75in,
  footskip=20pt
]{geometry}
\usepackage{setspace}
\usepackage[super,sort&compress]{natbib}
\AtBeginDocument{\setcitestyle{citesep={,}}}
\usepackage{microtype}

\usepackage{amsmath}
\usepackage{graphicx}
\usepackage{float}
\usepackage{booktabs}
\usepackage{tabularx}
\usepackage{array}
\usepackage{multirow}
\usepackage{threeparttable}
\usepackage{caption}
\usepackage{titlesec}
\usepackage{hyperref}
\hypersetup{hidelinks}

\usepackage[dvipsnames]{xcolor}

\titleformat{\section}
  {\fontsize{14}{18}\selectfont\bfseries}{}{0pt}{}
\titleformat{\subsection}
  {\normalsize\bfseries}{}{0pt}{}
\titleformat{\subsubsection}
  {\normalsize\bfseries}{}{0pt}{}
\titlespacing*{\section}{0pt}{28pt}{4pt}
\titlespacing*{\subsection}{0pt}{17pt}{5pt}
\titlespacing*{\subsubsection}{0pt}{14pt}{4pt}

\renewcommand{\figurename}{Fig.}
\newcolumntype{Y}{>{\raggedright\arraybackslash}X}

\begin{document}

\thispagestyle{empty}
\vspace*{-0.45in}
\begin{center}
  {\fontsize{17}{22}\selectfont\bfseries
  Contactless gait-derived features as candidate digital biomarkers of depression severity in adolescents\par}
\end{center}

\vspace{0.15in}
\begin{center}
  \begin{center}
Shuai Liu$^{1,2,\dagger}$,
Xiang Li$^{4,\dagger}$,
Guangxin Dai$^{1,\dagger}$,
Mengyuan Yu$^{3,\dagger}$,
Wenkong Wang$^{1,2}$,
Hanchi Liu$^{1,2}$,\
Jiahui Niu$^{1,2}$,
Yujie Nie$^{1,2}$,
Yumei Wang$^{3,*}$,
Xin Ma$^{1,2,*}$,
and Lizhou Fan$^{4,*}$
\par
\end{center}

\vspace{0.12in}

{\fontsize{9.7}{10.7}\selectfont
\setlength{\parindent}{0pt}
\setlength{\parskip}{0pt}
$^{1}$School of Control Science and Engineering, Shandong University, Jinan, Shandong, China\par
$^{2}$Engineering Research Center of Intelligent Unmanned System, Ministry of Education, China\par
$^{3}$Department of Psychology, Shandong Provincial Hospital Affiliated to Shandong First Medical University, Jinan, Shandong, China\par
$^{4}$Department of Psychiatry, The Chinese University of Hong Kong, Shatin, Hong Kong SAR, China\par

\vspace{0.10in}
$^{\dagger}$These authors contributed equally: Shuai Liu, Xiang Li, \textcolor{blue!50!black}{Guangxin Dai,} and Mengyuan Yu.\par
$^{*}$Correspondence should be addressed to Yumei Wang (wangym@bjmu.edu.cn), Xin Ma (maxin@sdu.edu.cn), or Lizhou Fan (leofan@cuhk.edu.hk)
}
\end{center}

\clearpage

\noindent
{\fontsize{14}{18}\selectfont\bfseries Abstract:}\par
\vspace{8pt}

Timely assessment of depression severity in adolescents is essential for guiding clinical care, yet conventional evaluations rely on episodic clinical assessments and symptom rating scales. Contactless gait analysis offers a potentially scalable approach to capturing objective behavioral features associated with depression. Here, we introduce the Long-Distance 3D Adolescent Depression Gait Dataset (LD3D-ADG), comprising depth-sensor gait recordings from 119 adolescents at a single center, including 39 healthy controls, 41 with mild depression, and 39 with moderate-to-severe depression. We developed UMLPA-GCN, a graph convolutional network that models spatiotemporal gait dynamics across three anatomically defined skeletal partitions. In participant-independent five-fold cross-validation, UMLPA-GCN achieved 75.63\% classification accuracy, a macro-F1 score of 75.52\%, and a macro-area under the receiver operating characteristic curve of 0.839 (95\% confidence interval, 0.779--0.899). Class-specific recall was 87.18\%, 63.41\%, and 76.92\% for healthy controls, mild depression, and moderate-to-severe depression, respectively. At the participant level, reduced arm swing, slower pelvic movement, restricted joint range of motion, and greater arm-swing asymmetry were identified as leading potential digital biomarkers contributing to depression severity assessment, associated with higher 17-item Hamilton Depression Rating Scale scores. These findings provide proof-of-concept evidence for contactless digital depression severity stratification in adolescents. Further external and prospective validation is needed to establish their utility for longitudinal monitoring and timely clinical assessment.

% Severity assessment of adolescent depression relies on clinician-administered rating scales, which are difficult to repeat between visits. Here we present the Long-Distance 3D Adolescent Depression Gait Dataset (LD3D-ADG), a single-center cohort of 119 adolescents (39 healthy controls, 41 with mild depression, and 39 with moderate-to-severe depression) with depth-sensor gait recordings. We also develop UMLPA-GCN, which divides the skeleton into three anatomically defined body partitions before CTR-GCN-based spatiotemporal modeling. In participant-independent five-fold cross-validation, UMLPA-GCN achieved 75.63\% accuracy and a macro-F1 of 75.52\%, with a macro-AUC of 0.839 (95\% CI 0.779--0.899); recalls were 87.18\%, 63.41\%, and 76.92\% for healthy controls, mild depression, and moderate-to-severe depression, respectively. At the participant level, reduced arm swing, reduced pelvis path speed, reduced joint range of motion, and greater arm-swing asymmetry were associated with higher 17-item Hamilton Depression Rating Scale (HAMD-17) scores. These results constitute proof-of-concept evidence for low-burden severity stratification rather than diagnosis.

\vspace{0.45cm}

\newpage

\section{Introduction}

Adolescent depression is a disabling psychiatric condition that contributes substantially to mental health burden during the transition from childhood to adulthood \cite{Yang2024AdolescentBurden,Kieling2024YouthMentalBurden}. Population studies have also reported increasing rates of depressive episodes among adolescents and young adults \cite{Mojtabai2016,Lu2019AdolescentDepression}. Depression during this developmental period can affect academic, social, and daily functioning and is associated with greater healthcare use and adverse mental health outcomes later in life \cite{Malhi2018Depression,Korczak2023AdolescentDepression,Wright2016,Johnson2018,Bernaras2019}. Methods that quantify clinically relevant behavioral indicators can complement symptom interviews and support repeated assessment between visits.

Depression severity is commonly assessed using clinician-rated instruments such as the 17-item Hamilton Depression Rating Scale (HAMD-17) \cite{Hamilton1960}. These instruments remain essential, but their scores depend on the clinical encounter, patient reporting, and rater judgment. They are also difficult to obtain outside scheduled visits, which limits continuous monitoring between appointments \cite{Mao2023AutoDepressionDiagnosis}. Digital biomarkers derived from behavior or physiology can provide complementary, repeatable measurements, although their clinical validity remains to be established and their measurement procedures require careful standardization \cite{Zierer2024DigitalBiomarkers}. Prior work has examined electroencephalography \cite{Yasin2021EEGReview,Safayari2021EEGDLReview}, facial expression \cite{Fu2023FacialExpression,Guo2021FacialExpression}, speech \cite{Cummins2015Speech,Liu2024SpeechMeta}, and eye movement \cite{Zhang2022EyeMovement,Gao2023EyeTracking}. Some of these modalities require specialized equipment, active engagement, or identifiable audio-visual recordings, which constrain repeated or scalable use \cite{GarciaCeja2018,Reinertsen2018}.

Gait is a candidate psychomotor digital biomarker because walking is naturally occurring, non-invasive, and jointly reflects postural control, motor vigor, and affect-related motor behavior \cite{BelvederiMurri2020GaitReview,Elkjaer2022MotorAlterations}. Depth sensors capture gait as three-dimensional joint coordinates without retaining full-color video, although skeletal data remain sensitive and require appropriate governance \cite{Zhao2019SeeMentalState}. Depression-related gait findings include reduced gait velocity, reduced arm swing, altered trunk posture, and diminished head movement \cite{Lemke2000SpatiotemporalGait,Michalak2009Embodiment,Wang2021GaitContribution,Adolph2021GaitMood}. Motor abnormalities may emerge during development, so findings from adult populations cannot be assumed to generalize to adolescents \cite{Damme2022MotorAdolescence,Chan2022ChildGaitMood}.

Earlier automated approaches relied on handcrafted temporal, spectral, and geometric gait features \cite{Yuan2019HilbertHuang,Lu2021NewSkeletalRepresentation,Fang2019GaitAbnormality,Wang2021GaitFramework}. These were supplanted by recurrent, temporal-convolutional, multimodal, and augmentation-based deep models that learn directly from gait sequences \cite{Shao2022MultimodalGait,Li2021SimpleAction,Yang2022DataAugmentation,Lu2022MultimediaGait}. Graph-based methods have become increasingly prominent in this line of work: recent benchmarks treat the skeleton as a graph of joints \cite{Liu2024Benchmark,Li2025STMG}, and representative architectures such as ST-GCN \cite{Yan2018STGCN} and CTR-GCN \cite{Chen2021CTRGCN} have become de facto backbones for skeleton sequence modeling. These models were primarily designed for general action recognition and do not explicitly represent the clinically motivated distinction between upper-body posture, trunk and arm movement, and lower-limb locomotion.

To address this gap, we developed the Long-Distance 3D Adolescent Depression Gait Dataset (LD3D-ADG), comprising long-distance three-dimensional skeleton gait recordings from 119 adolescents, and UMLPA-GCN, an Upper--Middle--Lower Partition-Aware Graph Convolutional Network. UMLPA-GCN divides each skeleton sequence into anatomically defined upper, middle, and lower partitions, refines each partition with positional encoding and self-attention, fuses the partitions, and passes the result to a CTR-GCN backbone \cite{Chen2021CTRGCN}.

We evaluated whether this structured prior improved three-class participant-level classification of healthy controls, mild depression, and moderate-to-severe depression (study-defined HAMD-17 strata \cite{Zimmerman2013HAMDSeverity}), and analyzed interpretable participant-level kinematic features to examine whether skeletal gait patterns are associated with continuous HAMD-17 total scores. The study is a proof-of-concept analysis of severity-associated gait representation rather than a replacement for clinical diagnosis.

\begin{figure}[htbp]
\centering
\includegraphics[width=0.98\linewidth]{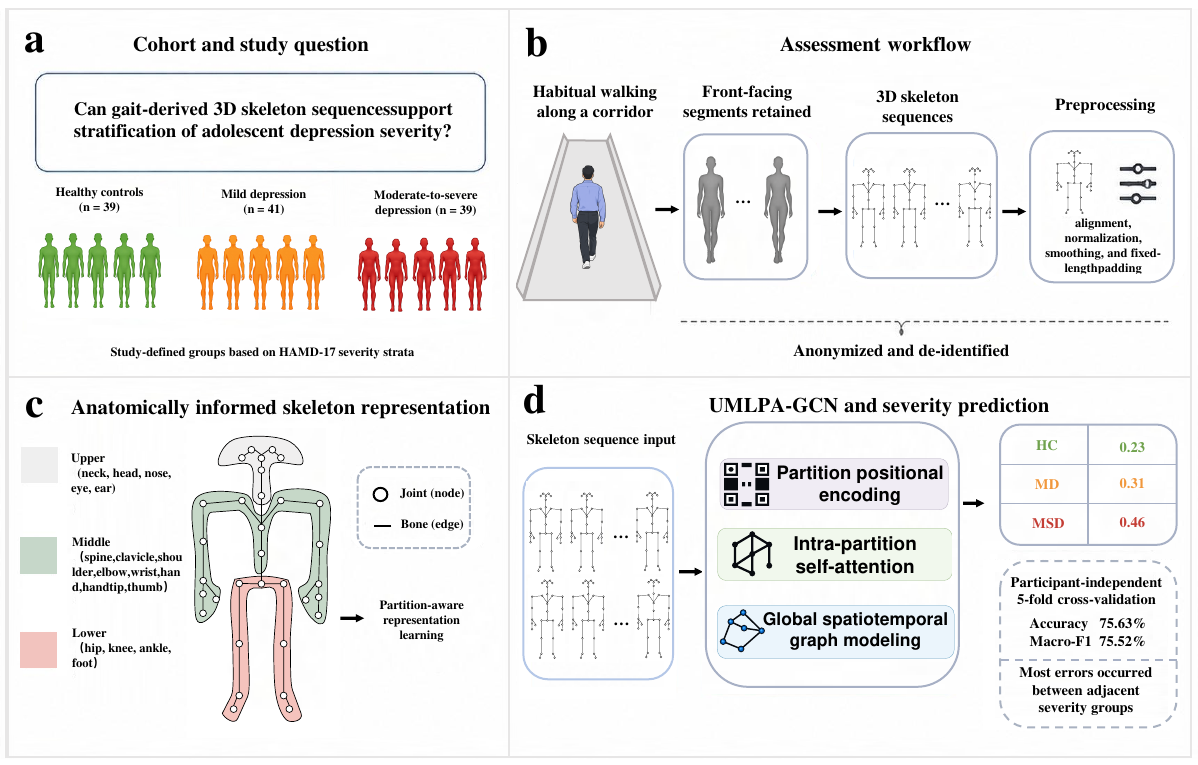}
\caption{Overview of the proposed framework for adolescent depression severity assessment. (a) Cohort definition and study population based on HAMD-17 severity strata. (b) Assessment workflow from habitual walking acquisition to frontal-view skeleton sequence preprocessing. (c) Anatomically informed skeleton representation with upper, middle, and lower body regions. (d) UMLPA-GCN-based feature learning and depression severity prediction.}
\label{fig:workflow}
\end{figure}

\section{Results}

\subsection{Participant characteristics}

The study included 119 participants, comprising 39 healthy controls, 41 participants with mild depression, and 39 participants with moderate-to-severe depression. The three groups were similar in size and had broadly comparable sex distributions. Healthy controls were older than both depression groups on average (Table~\ref{tab:datasetstats}).

\begin{table}[htbp]
\centering
\begin{threeparttable}
\caption{\textbf{Participant demographic and clinical characteristics.}}
\label{tab:datasetstats}
\begin{tabular}{lccc}
\toprule
Variable & Healthy controls & Mild depression & Moderate-to-severe depression \\
\midrule
Participants, no. & 39 & 41 & 39 \\
Sex, male/female & 15/24 & 13/28 & 12/27 \\
Age, years & $17.11 \pm 2.16$ & $16.34 \pm 2.02$ & $16.54 \pm 2.09$ \\
HAMD-17 total score & $1.95 \pm 2.29$ & $12.00 \pm 1.73$ & $23.38 \pm 2.54$ \\
\bottomrule
\end{tabular}
\begin{tablenotes}[flushleft]\footnotesize
\item Values are mean $\pm$ standard deviation unless otherwise stated. HAMD-17, 17-item Hamilton Depression Rating Scale.
\end{tablenotes}
\end{threeparttable}
\end{table}

\subsection{Primary classification performance}

Across pooled out-of-fold participant predictions, UMLPA-GCN achieved an accuracy of 0.7563 and a macro-F1 of 0.7552, the highest values among the seven evaluated skeleton-based models (Table~\ref{tab:overall_comparison}). The corresponding values for the CTR-GCN backbone were 0.7227 and 0.7225. These differences represent absolute increases of 3.36 percentage points in accuracy and 3.27 percentage points in macro-F1. Because no uncertainty intervals or paired comparisons are available, these differences should not be interpreted as evidence of superiority.

\begin{table}[htbp]
\centering
\begin{threeparttable}
\caption{\textbf{Participant-level performance under five-fold cross-validation.}}
\label{tab:overall_comparison}
\begin{tabular}{lcccc}
\toprule
Method & Accuracy & Macro-precision & Macro-recall & Macro-F1 \\
\midrule
ST-GCN \cite{Yan2018STGCN} & 0.6891 & 0.6889 & 0.6900 & 0.6892 \\
2s-AGCN \cite{Shi2019AGCN} & 0.6975 & 0.6973 & 0.6986 & 0.6976 \\
MS-G3D \cite{Liu2020MSG3D} & 0.7059 & 0.7054 & 0.7067 & 0.7058 \\
HD-GCN \cite{Lee2023HDGCN} & 0.7143 & 0.7145 & 0.7152 & 0.7143 \\
InfoGCN \cite{Chi2022InfoGCN} & 0.7227 & 0.7226 & 0.7230 & 0.7221 \\
CTR-GCN \cite{Chen2021CTRGCN} & 0.7227 & 0.7229 & 0.7234 & 0.7225 \\
\textbf{UMLPA-GCN} & \textbf{0.7563} & \textbf{0.7541} & \textbf{0.7584} & \textbf{0.7552} \\
\bottomrule
\end{tabular}
\begin{tablenotes}[flushleft]\footnotesize
\item Metrics are computed from pooled out-of-fold predictions over all 119 participants (see Fig.~\ref{fig:confusion_matrix}a).
\end{tablenotes}
\end{threeparttable}
\end{table}

\subsection{Class-specific error pattern}

\begin{figure}[htbp]
\centering
\includegraphics[width=1\linewidth]{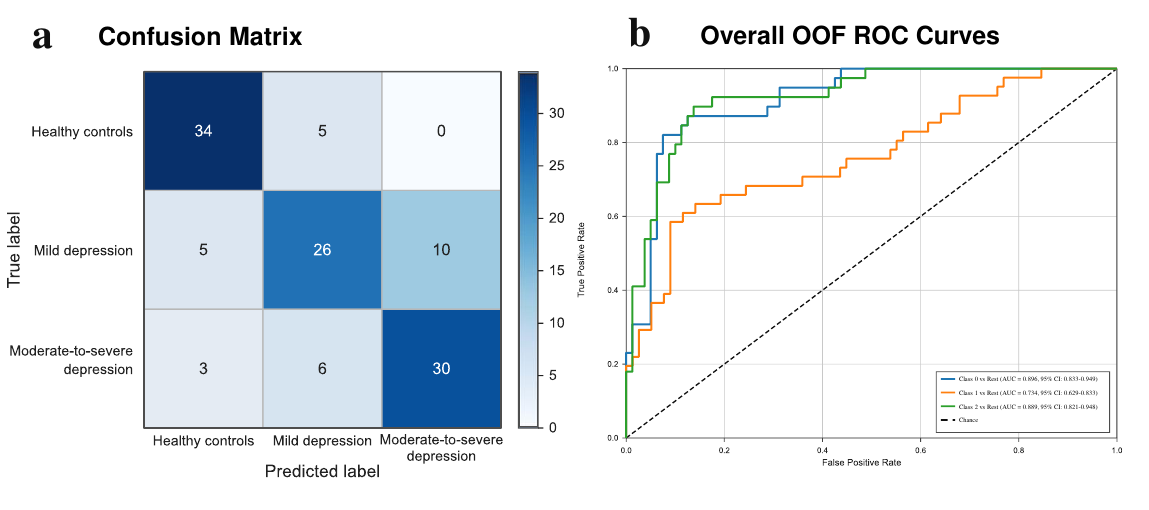}
\caption{\textbf{Classification performance of UMLPA-GCN on the LD3D-ADG dataset.} (a) Pooled out-of-fold confusion matrix: rows indicate study-defined HAMD-17 groups and columns indicate predicted groups; cell entries show participant counts ($n=119$). (b) Overall out-of-fold (OOF) ROC curves for one-vs-rest adolescent depression severity classification. HC, healthy controls; MD, mild depression; MSD, moderate-to-severe depression.}
\label{fig:confusion_matrix}
\end{figure}

The pooled out-of-fold confusion matrix contained 90 correct predictions among 119 participants (Fig.~\ref{fig:confusion_matrix}). Recall was 87.18\% for healthy controls, 63.41\% for mild depression, and 76.92\% for moderate-to-severe depression. Ten of 41 participants with mild depression were classified as moderate-to-severe, while six of 39 participants with moderate-to-severe depression were classified as mild. No healthy control was classified as moderate-to-severe depression. The concentration of errors between adjacent severity categories may reflect the fact that HAMD-17 strata impose discrete boundaries on a continuous symptom construct. The lower recall for mild depression matters because this group may be most relevant to early identification. The observed error pattern therefore supports broad severity-associated stratification more strongly than precise clinical staging.

\subsection{Anatomical partition ablation}

Accuracy increased as the partition scheme went from a single whole-body partition to two partitions and then to the three-partition upper, middle, and lower scheme (Table~\ref{tab:partition_ablation}). The three-partition model achieved an accuracy of 0.7563 compared with 0.7311 for the single-partition model. Within this single dataset, finer anatomical partitioning was associated with higher accuracy.

\begin{table}[htbp]
\centering
\caption{\textbf{Ablation of anatomical partition granularity.}}
\label{tab:partition_ablation}
\begin{threeparttable}
\begin{tabular}{lcccc}
\toprule
Partition scheme & Accuracy & Macro-precision & Macro-recall & Macro-F1 \\
\midrule
Single partition & 0.7311 & 0.7320 & 0.7290 & 0.7292 \\
Two partitions & 0.7479 & 0.7482 & 0.7482 & 0.7455 \\
Three partitions & \textbf{0.7563} & \textbf{0.7541} & \textbf{0.7584} & \textbf{0.7552} \\
\bottomrule
\end{tabular}
\begin{tablenotes}[flushleft]\footnotesize
\item "Single partition" uses the whole skeleton as one graph; "two partitions" divides the skeleton into upper+middle and lower regions; "three partitions" splits it into upper (head and neck), middle (torso and arms), and lower (legs and feet) anatomical regions.
\end{tablenotes}
\end{threeparttable}
\end{table}

\subsection{Component ablation}

Adding the upper--middle--lower partition module (UMLPM) to CTR-GCN increased the accuracy point estimate from 0.7227 to 0.7311 (Table~\ref{tab:component_ablation}). Adding partition positional encoding (PPE) or intra-partition self-attention (IPSA) yielded further increases, to 0.7395 and 0.7479, respectively. The full UMLPA-GCN produced the highest point estimates, consistent with contributions from anatomical partitioning, partition-aware encoding, and within-partition refinement; because the ablation adds components sequentially, it does not isolate pairwise interactions.

\begin{table}[htbp]
\centering
\caption{\textbf{Ablation of UMLPA-GCN components.}}
\label{tab:component_ablation}
\begin{tabular}{lcccc}
\toprule
Variant & Accuracy & Macro-precision & Macro-recall & Macro-F1 \\
\midrule
CTR-GCN & 0.7227 & 0.7229 & 0.7234 & 0.7225 \\
CTR-GCN + UMLPM & 0.7311 & 0.7335 & 0.7302 & 0.7306 \\
CTR-GCN + UMLPM + PPE & 0.7395 & 0.7389 & 0.7421 & 0.7365 \\
CTR-GCN + UMLPM + IPSA & 0.7479 & 0.7394 & 0.7444 & 0.7388 \\
\textbf{Full UMLPA-GCN} & \textbf{0.7563} & \textbf{0.7541} & \textbf{0.7584} & \textbf{0.7552} \\
\bottomrule
\end{tabular}
\end{table}

\subsection{Threshold-independent discrimination}

Fig.~\ref{fig:confusion_matrix}(b) shows the one-vs-rest ROC curves constructed from pooled out-of-fold prediction probabilities, which characterize class-wise discrimination independently of a fixed decision threshold. The macro-averaged out-of-fold AUC was 0.839 (95\% confidence interval: 0.779--0.899), indicating good overall discrimination across the three severity classes in this single-cohort, internally cross-validated setting. At the class level, the healthy-control group achieved an AUC of 0.896 (95\% confidence interval: 0.833--0.949) and the moderate-to-severe group an AUC of 0.889 (95\% confidence interval: 0.821--0.948), whereas the intermediate mild-depression group showed a lower AUC of 0.734 (95\% confidence interval: 0.629--0.833). This pattern is consistent with the confusion matrix: in this cohort, the model discriminated the healthy-control and moderate-to-severe groups more accurately, whereas the intermediate mild group overlapped more with the adjacent severity groups.

\subsection{Participant-level kinematic correlates of depression severity}
\label{subsec:clinical-consistency}

To examine whether the raw skeletal gait data contain clinically interpretable, severity-related motor information independent of the classifier, we extracted interpretable kinematic features at the participant level. The retained frontal-view sequences were aggregated within each participant, yielding one value per participant and per feature. The feature set included pelvis path speed, head movement, trunk posture, arm swing amplitude and asymmetry, lower-limb movement, and joint ranges of motion (ROM). Associations with continuous HAMD-17 total scores were evaluated using Spearman's rank correlation. Nominal $P$-values were adjusted with the Benjamini--Hochberg false discovery rate (BH-FDR) procedure, and Pearson correlations with bootstrap 95\% confidence intervals and exploratory $\eta^2$ values were computed descriptively. Within-group sensitivity analyses were performed within each severity group.

\begin{figure}[htbp]
\centering
\includegraphics[width=0.6\linewidth]{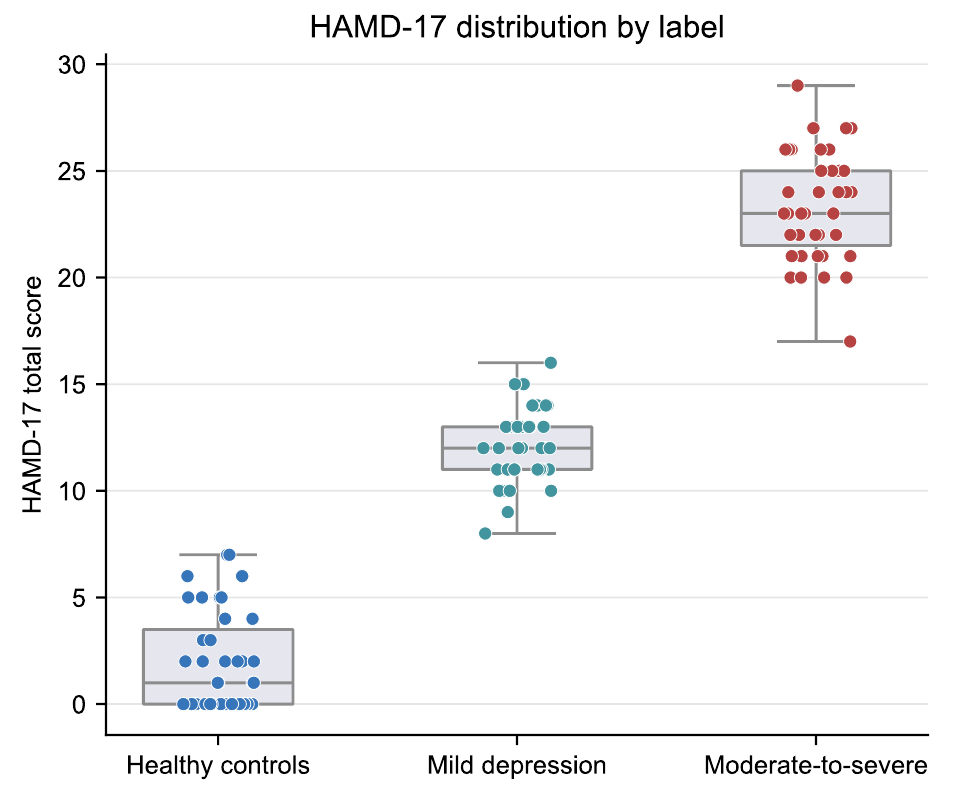}
\caption{Distribution of HAMD-17 total scores across healthy controls (HC), mild depression (MD), and moderate-to-severe depression (MSD). The observed separation is consistent with the predefined HAMD-17-based severity cutoffs.}
\label{fig:hamd_distribution}
\end{figure}

The analysis included all 119 participants: 39 healthy controls (HC), 41 with mild depression (MD), and 39 with moderate-to-severe depression (MSD). Fig.~\ref{fig:hamd_distribution} shows the HAMD-17 total score distribution across the three groups, and Supplementary Fig.~S1 shows the exploratory between-group kinematic effect-size ranking. Left and right arm swing range, mean elbow ROM, and pelvis path speed showed relatively large $\eta^2$ values.

Supplementary Fig.~S2 shows the group-wise distributions of the representative kinematic features and the standardized group-mean profile. From HC to MD and MSD, pelvis path speed, left arm swing range, and mean elbow ROM generally decreased, whereas arm swing asymmetry increased. Left arm swing range decreased from 0.873 in HC to 0.589 in MD and 0.327 in MSD, the largest exploratory between-group effect among the examined features ($\eta^2=0.493$). Mean elbow ROM decreased from $43.6^\circ$ to $41.0^\circ$ and $22.3^\circ$ across the three groups ($\eta^2=0.375$), whereas pelvis path speed (body lengths/s) decreased from 3.12 to 2.92 and 2.52 ($\eta^2=0.330$). Arm swing asymmetry increased from 21.2\% in HC to 45.0\% in MD and 51.0\% in MSD ($\eta^2=0.158$).

\begin{figure}[htbp]
\centering
\includegraphics[width=\textwidth]{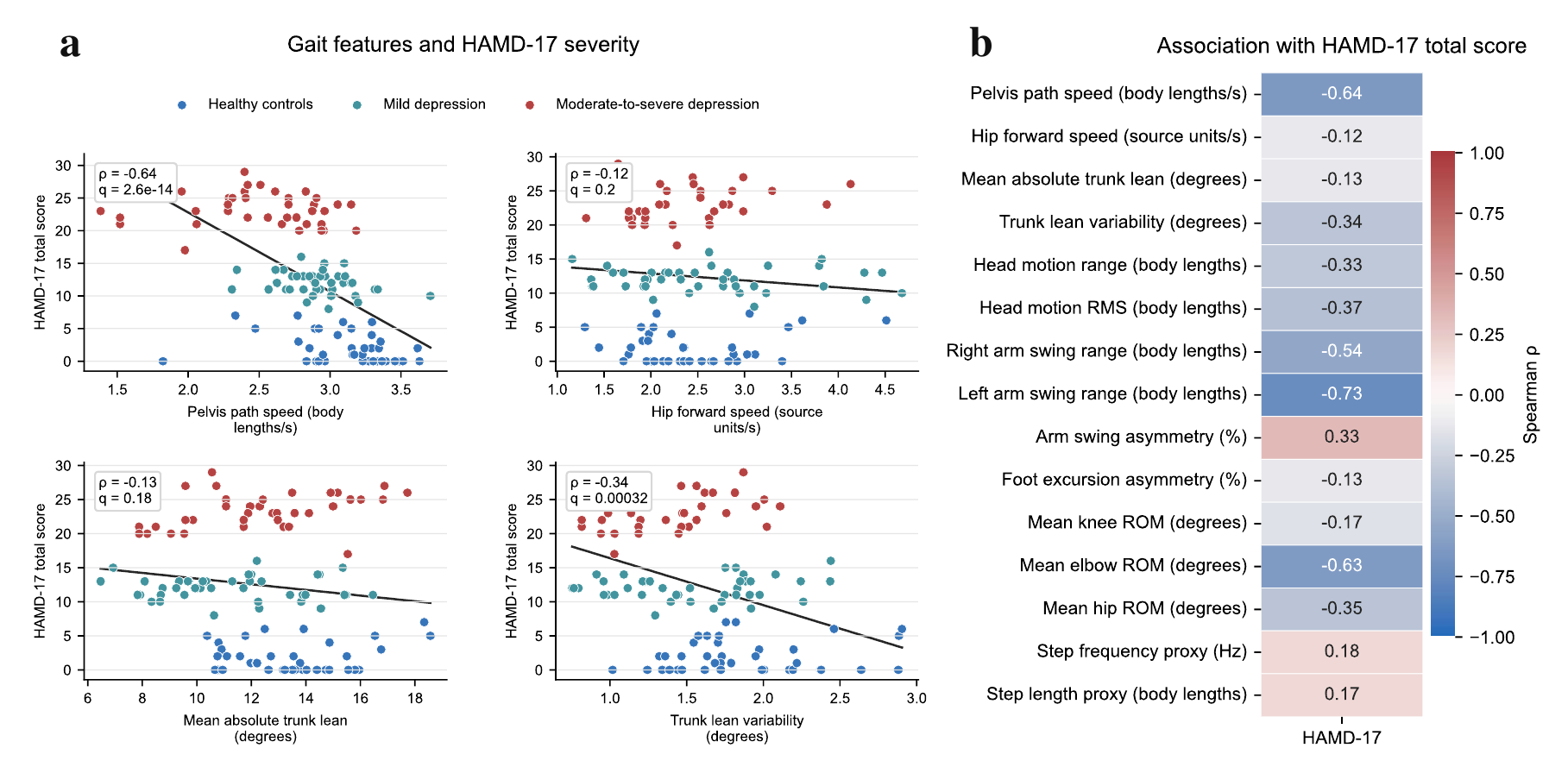}
\caption{Participant-level associations between representative kinematic features and HAMD-17 total score. (a) Scatter plots showing HAMD-17 total score against pelvis path speed (body lengths/s), hip forward speed (source units/s), mean absolute trunk lean (degrees), and trunk lean variability (degrees). Each point represents one participant ($n=119$), and solid lines indicate fitted linear trends for visualization only; statistical inference was based on Spearman's rank correlation. (b) Spearman correlation coefficients between HAMD-17 total score and the 15 predefined kinematic features.}
\label{fig:kinematic_corr}
\end{figure}

In the pooled cohort of 119 participants, several features remained significant after BH-FDR correction (Table~\ref{tab:kinematic_correlation}). Fig.~\ref{fig:kinematic_corr} displays participant-level scatter plots for pelvis path speed, hip forward speed, mean absolute trunk lean, and trunk lean variability against HAMD-17 total score, together with a Spearman correlation summary for all 15 predefined kinematic features. Left arm swing range showed the strongest negative association with HAMD-17 total score ($\rho=-0.727$, $q=1.28\times10^{-19}$). Pelvis path speed ($\rho=-0.642$, $q=2.56\times10^{-14}$) and mean elbow ROM ($\rho=-0.631$, $q=7.19\times10^{-14}$) also showed negative associations. Right arm swing range was negatively associated with symptom severity ($\rho=-0.545$, $q=5.57\times10^{-10}$). Head movement root mean square (RMS), hip ROM, and trunk lean variability showed moderate negative correlations. Arm swing asymmetry was positively correlated with symptom severity ($\rho=0.327$, $q=5.08\times10^{-4}$). Cadence and step-length proxy measures showed weaker associations and did not remain significant after BH-FDR correction.

\begin{table}[htbp]
\centering
\caption{Participant-level associations between interpretable kinematic features and HAMD-17 total score.}
\label{tab:kinematic_correlation}
\resizebox{\textwidth}{!}{%
\begin{tabular}{lccccc}
\toprule
\textbf{Kinematic feature} &
\textbf{Spearman $\boldsymbol{\rho}$} &
\textbf{\textit{P}-value} &
\textbf{BH-\textit{q}} &
\textbf{Pearson $\boldsymbol{r}$} &
\textbf{95\% CI} \\
\midrule
Left arm swing range (body lengths)
& $-0.727$
& $8.55 \times 10^{-21}$
& $1.28 \times 10^{-19}$
& $-0.721$
& $[-0.792,\,-0.639]$ \\

Pelvis path speed (body lengths/s)
& $-0.642$
& $3.41 \times 10^{-15}$
& $2.56 \times 10^{-14}$
& $-0.586$
& $[-0.700,\,-0.451]$ \\

Mean elbow ROM (degrees)
& $-0.631$
& $1.44 \times 10^{-14}$
& $7.19 \times 10^{-14}$
& $-0.578$
& $[-0.708,\,-0.445]$ \\

Right arm swing range (body lengths)
& $-0.545$
& $1.49 \times 10^{-10}$
& $5.57 \times 10^{-10}$
& $-0.584$
& $[-0.689,\,-0.452]$ \\

Head movement RMS
& $-0.370$
& $3.39 \times 10^{-5}$
& $1.02 \times 10^{-4}$
& $-0.302$
& $[-0.476,\,-0.147]$ \\

Hip ROM (degrees)
& $-0.348$
& $1.03 \times 10^{-4}$
& $2.58 \times 10^{-4}$
& $-0.327$
& $[-0.490,\,-0.159]$ \\

Trunk lean variability (degrees)
& $-0.341$
& $1.51 \times 10^{-4}$
& $3.23 \times 10^{-4}$
& $-0.372$
& $[-0.503,\,-0.220]$ \\

Arm swing asymmetry (\%)
& $+0.327$
& $2.85 \times 10^{-4}$
& $5.08 \times 10^{-4}$
& $+0.346$
& $[+0.207,\,+0.484]$ \\

Cadence proxy
& $+0.176$
& $0.056$
& $0.082$
& ---
& --- \\

Step length proxy
& $+0.173$
& $0.060$
& $0.082$
& ---
& --- \\
\bottomrule
\end{tabular}%
}
\vspace{1mm}
\begin{minipage}{0.97\textwidth}
\footnotesize
\textit{Note:} Associations were calculated at the participant level ($n=119$). Spearman's rank correlation coefficients were used as the primary association measure. $P$-values were adjusted using the Benjamini--Hochberg false discovery rate (BH-FDR) procedure. Pearson correlation coefficients and bootstrap 95\% confidence intervals (CIs) are reported for selected representative features. Positive values indicate that the kinematic feature increases with HAMD-17 total score, whereas negative values indicate a decrease with increasing depression severity. ROM, range of motion; RMS, root mean square. 
\end{minipage}
\end{table}

We then performed within-group sensitivity analyses to assess whether the pooled associations held within each severity group. In the HC group, left arm swing range remained negatively associated with HAMD-17 total score ($\rho=-0.488$, $P=0.0016$, within-group BH-FDR-adjusted $q=0.0245$). Pelvis path speed showed a weaker association ($\rho=-0.419$, $P=0.0080$, $q=0.0596$). No predefined feature reached an uncorrected significance level of $P<0.05$ in the MD group. In the MSD group, mean absolute trunk lean, head movement RMS, and foot movement asymmetry were associated with HAMD-17 total score. Fewer within-group associations reached significance than in the pooled cohort, a pattern that likely reflects the restricted HAMD-17 range within each group and the limited within-group sample sizes.

Higher depression severity was primarily associated with reduced movement amplitude, reduced joint mobility, and reduced bilateral coordination, whereas cadence showed weaker associations. The identified features span the upper limbs, trunk/pelvis, and lower limbs, the region-specific pattern that motivated the upper--middle--lower partition design. These analyses describe the gait data itself; they do not demonstrate that individual network decisions were driven by these handcrafted features.

\section{Discussion}

This study evaluated whether anatomically structured skeleton representations capture gait patterns associated with adolescent depression severity. UMLPA-GCN yielded an accuracy of 75.63\% and macro-F1 of 75.52\% from pooled participant-level out-of-fold predictions. The model achieved its highest point estimates with three body partitions and all partition-refinement components. The confusion matrix showed better separation of healthy controls than differentiation between adjacent depression-severity categories, with the mild-depression group having the lowest recall.

The anatomical partition prior was motivated by the psychomotor expression of depression: prior observational work has associated depression with slower locomotion, reduced arm swing, postural flexion, and diminished head movement \cite{Lemke2000SpatiotemporalGait,Michalak2009Embodiment,Wang2021GaitContribution}, changes that are distributed unevenly across the body. The ablation results are consistent with this hypothesis, but these results do not establish that the model has identified specific pathophysiological markers; no region-level attribution or occlusion analysis was performed.

The participant-level kinematic analysis was performed on the raw gait data, independently of the classifier. Higher HAMD-17 total scores were associated with the same pattern reported in the Results: reduced movement amplitude and joint mobility and greater arm-swing asymmetry. The associated features involve the upper limbs, trunk/pelvis, and lower limbs, a region-specific pattern consistent with the motivation for the upper--middle--lower partition prior.

A skeleton-based workflow allows repeated assessment. Three-dimensional joint coordinates omit appearance information, and their identifiability relative to full-color video has not been tested here. Depth sensing requires no wearable device and no active questionnaire. These properties could support low-burden measurement in clinical follow-up or research settings. Nevertheless, skeletal trajectories remain sensitive behavioral data, and the present system was evaluated only in a controlled hospital corridor. The findings support further biomarker development rather than deployment for diagnosis, triage, or treatment decisions.

Healthy controls were identified with comparatively high recall, whereas mild and moderate-to-severe depression were more frequently confused. This may reflect overlapping psychomotor manifestations, noise around HAMD-17 thresholds, or insufficient sample size. On the observed error pattern, gait alone is better suited to complementing symptom assessment than to assigning precise severity categories. Future studies should examine continuous HAMD-17 prediction, ordinal modeling, clinically meaningful change, and multimodal integration rather than relying only on three discrete classes.

Several limitations constrain interpretation. The study was small, single-center, and cross-sectional, and it was evaluated by internal cross-validation without an independent external cohort. Healthy controls were on average older than both depression groups, and because group assignment was derived from HAMD-17 strata, part of this difference may be endogenous to the labeling scheme. Age is a known determinant of gait, and age-matched or covariate-adjusted analyses have not been performed, so the reported accuracy may be inflated by age-related gait differences and by the internal, single-cohort validation design, and should not be taken as an estimate of external generalization. Severity labels were based on study-defined HAMD-17 strata at one time point and do not capture longitudinal symptom change, inter-rater variability, or the full clinical heterogeneity of adolescent depression. Only frontal-view skeletal segments from a controlled corridor were retained, which limits evidence for robustness across viewpoints, sites, clothing, and environments. Uncertainty estimates, calibration, test-retest reliability, subgroup performance, and formal comparisons between models were not available. The model used gait alone and was not evaluated against clinician judgment, established screening instruments, or complementary behavioral modalities.

Future validation should prioritize multicenter and longitudinal cohorts, participant-level preregistered analysis, age and sex matching, transparent handling of missing or low-confidence joints, temporal splitting to guard against within-participant autocorrelation, and external testing across sensors and environments. Clinical studies should evaluate whether gait features add information beyond demographic variables and standard symptom scales, whether changes track treatment response, and whether outputs can be presented with calibrated uncertainty. Under these conditions, anatomically structured skeleton modeling can contribute to multimodal, privacy-conscious assessment workflows.

\section{Methods}

\subsection{Study design and ethics}

We conducted a single-center, cross-sectional study to develop and internally evaluate a skeleton-based model for three-class classification of study-defined adolescent depression severity. Participants were recruited from the Clinical Psychology Ward of Shandong Provincial Hospital Affiliated to Shandong First Medical University.

The study was approved by the Ethics Committee of Shandong Provincial Hospital Affiliated to Shandong First Medical University and was conducted in accordance with the Declaration of Helsinki. Written informed consent was obtained from participants and their legal guardians before study procedures. Participants received compensation for participation. Ethics approval number: R202411120385.

\subsection{Participants and clinical labels}

The LD3D-ADG cohort included 119 participants: 39 healthy controls, 41 participants with mild depression, and 39 participants with moderate-to-severe depression. Participants in the depression groups met DSM-5 criteria for major depressive disorder, were experiencing a current depressive episode, had received no treatment during the preceding 14 days, and had sufficient education and comprehension to complete study procedures. Exclusion criteria included current or previous major psychiatric disorders other than depression, serious organic disease, participation in an interventional clinical trial within the preceding three months, poor compliance, or investigator-determined unsuitability.

Healthy controls were recruited from a comparable source population and screened to exclude major psychiatric disorders and severe organic disease. Trained psychiatrists administered the HAMD-17 \cite{Hamilton1960}. Participants with scores of 0--7 were assigned to the healthy-control group, those with scores of 8--16 to the mild-depression group, and those with scores of 17 or higher to the moderate-to-severe group, following published empirically derived severity cutoffs for the 17-item scale \cite{Zimmerman2013HAMDSeverity}. Moderate and severe categories were combined because of limited subgroup size. The prediction target was the study-defined HAMD-17 category, not a stand-alone diagnosis. Participants were aged 14--20 years.

\subsection{Gait acquisition}

\begin{figure}[htbp]
\centering
\includegraphics[width=0.95\linewidth]{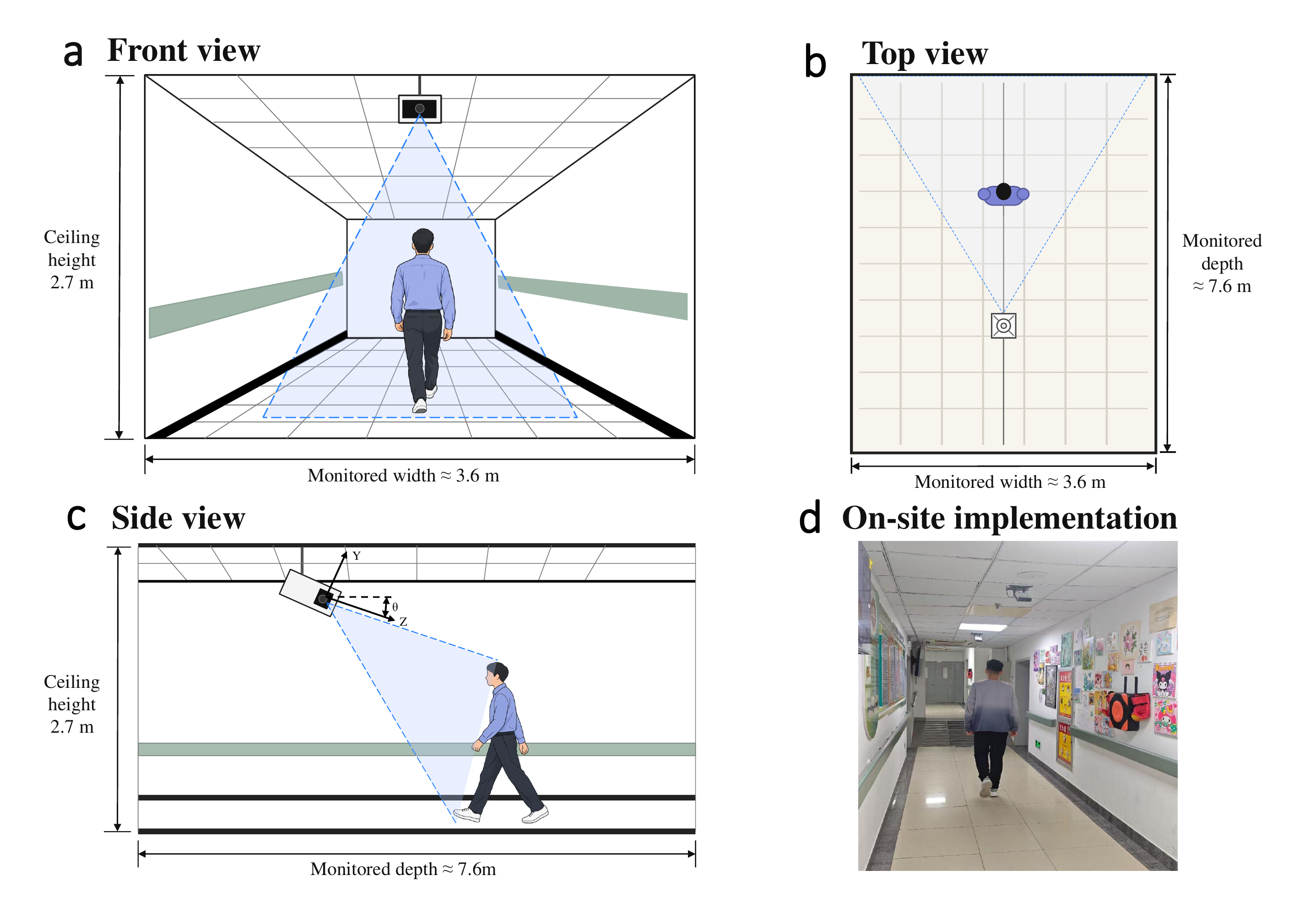}
\caption{\textbf{Walking paradigm and acquisition setup for long-distance gait data collection.} (a) Front view, (b) top view, and (c) side view of the corridor monitoring configuration with Azure Kinect DK sensors. (d) On-site implementation of the gait acquisition environment.}
\label{fig:paradigm}
\end{figure}

Participants completed two walking trials at a self-selected walking speed along a predefined 15.5-m corridor route. Two clinicians positioned at opposite ends of the corridor indicated turning points, monitored participant safety, and allowed rest between trials. Three Azure Kinect DK depth sensors recorded depth frames at $320 \times 288$ pixels and 30 frames per second. Three-dimensional joint coordinates were extracted with the Azure Kinect SDK. The acquisition arrangement is shown in Fig.~\ref{fig:paradigm}. The approximately 7.6-m monitored depth illustrated in Fig.~\ref{fig:paradigm}(b,c) represents the sensor observation range shown in the schematic rather than the full 15.5-m walking route.

\subsection{Skeleton preprocessing}

Raw sequences underwent view selection, coordinate alignment, root normalization, denoising, and temporal standardization. Participants walked in both directions. Only frontal-view segments were retained because back-view Azure Kinect skeletons were considered less reliable \cite{Lu2022MultimediaGait}. 

Because ceiling-mounted sensors were tilted relative to the walking plane, a line was fitted to the SpineBase coordinates $(z_i,y_i)$ across frames. The tilt angle $\theta$ was estimated as

\begin{equation}
\tan\theta =
\frac{n\sum_i z_i y_i-\left(\sum_i z_i\right)\left(\sum_i y_i\right)}
     {n\sum_i z_i^2-\left(\sum_i z_i\right)^2},
\label{eq:theta}
\end{equation}

where $n$ denotes the number of frames. Joint coordinates were then rotated around the $x$-axis:

\begin{equation}
\begin{pmatrix}
x' \\ y' \\ z'
\end{pmatrix}
=
\begin{pmatrix}
1 & 0 & 0 \\
0 & \cos\theta & -\sin\theta \\
0 & \sin\theta & \cos\theta
\end{pmatrix}
\begin{pmatrix}
x \\ y \\ z
\end{pmatrix}.
\label{eq:rotation}
\end{equation}

For each frame, SpineBase coordinates were subtracted from all joints to remove global translation. A Gaussian filter with a window length of five frames and a standard deviation of 1.0 frame was applied independently to each coordinate trajectory, consistent with prior gait preprocessing\cite{Wang2021GaitFramework}. Sequences longer than 50 frames were truncated to retain the first 50 frames, whereas shorter sequences were zero-padded to reach 50 frames.

\subsection{UMLPA-GCN architecture}

\begin{figure}[htbp]
\centering
\includegraphics[width=0.98\linewidth]{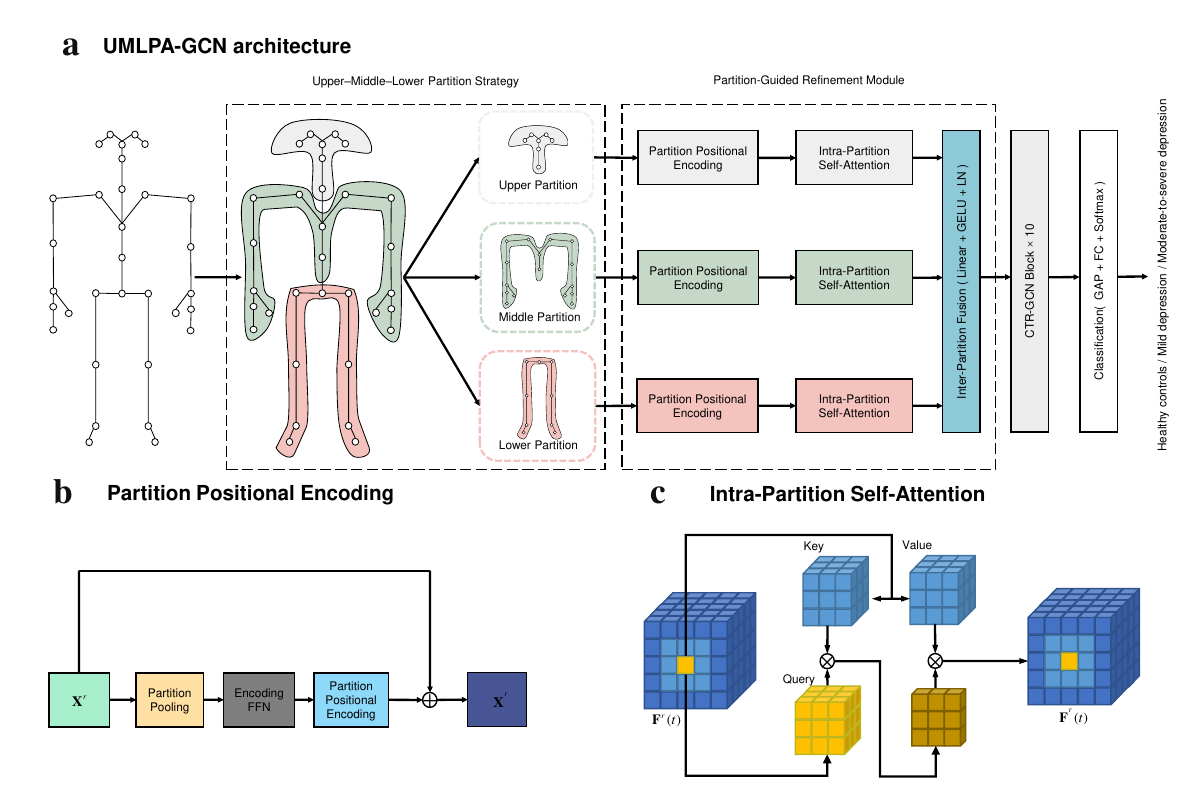}
\caption{\textbf{Detailed architecture of UMLPA-GCN.} (a) Overall UMLPA-GCN architecture, including upper--middle--lower body partitioning, partition-guided refinement, CTR-GCN backbone, and severity classification. (b) Partition Positional Encoding (PPE) module for injecting partition-aware information. (c) Intra-Partition Self-Attention (IPSA) module for adaptive joint refinement within each body region. Exact joint membership is provided in Supplementary Table~S2.}
\label{fig:framework}
\end{figure}

The overall study workflow is shown in Fig.~\ref{fig:workflow}, and the UMLPA-GCN architecture in Fig.~\ref{fig:framework}. An input skeleton sequence is divided into upper, middle, and lower anatomical partitions. Partition features are refined by Partition Positional Encoding (PPE), Intra-Partition Self-Attention (IPSA), and Inter-Partition Fusion (IPF). The fused representation is processed by ten CTR-GCN blocks \cite{Chen2021CTRGCN}, followed by global average pooling and a fully connected softmax classifier.

\subsubsection{Upper--middle--lower partition module}

Let $\mathbf{X}\in\mathbb{R}^{B\times C\times T\times V}$ denote the input, where $B$ is batch size, $C=3$ coordinate channels, $T$ is the number of frames, and $V=32$ joints. The joint set is divided into three non-overlapping partitions:

\begin{equation}
\mathcal{V}=\mathcal{V}^{u}\cup\mathcal{V}^{m}\cup\mathcal{V}^{l},
\qquad
\mathcal{V}^{a}\cap\mathcal{V}^{b}=\varnothing \quad (a\neq b).
\label{eq:uml_partition_set}
\end{equation}

For partition $r\in\{u,m,l\}$, a binary mask $\mathbf{M}^{r}\in\{0,1\}^{V}$ produces

\begin{equation}
\mathbf{X}^{r}=\mathbf{X}\odot\mathbf{M}^{r},
\qquad
\mathbf{M}^{u}+\mathbf{M}^{m}+\mathbf{M}^{l}=\mathbf{1},
\label{eq:uml_partition_mask}
\end{equation}

where broadcasting is applied over batch, channel, and time dimensions. The partition definition was motivated by reported depression-associated changes in head posture, trunk and arm movement, and lower-limb locomotion \cite{Lemke2000SpatiotemporalGait,Michalak2009Embodiment,Wang2021GaitContribution}. The exact mapping of the 32 Azure Kinect joints is given in Supplementary Table~S2.

\subsubsection{Partition positional encoding}

PPE generates a frame-level descriptor for each anatomical partition (Fig.~\ref{fig:framework}(b)). For partition $r$, with $V_r=|\mathcal{V}^{r}|$ joints,

\begin{equation}
\mathbf{d}^{r}(t)=\frac{1}{V_r}\sum_{v\in\mathcal{V}^{r}}
\mathbf{X}^{r}_{:,:,t,v},
\qquad
\mathbf{e}^{r}(t)=\mathrm{FFN}_{r}\left(\mathbf{d}^{r}(t)\right),
\label{eq:ppe_encode}
\end{equation}

where $\mathrm{FFN}_{r}$ is a two-layer partition-specific feed-forward network with Gaussian error linear unit activation. The encoding is broadcast to all joints in the partition and added residually:

\begin{equation}
\widehat{\mathbf{X}}^{r}_{:,:,t,v}
=
\mathbf{X}^{r}_{:,:,t,v}+\mathbf{e}^{r}(t),
\qquad
v\in\mathcal{V}^{r}.
\label{eq:ppe_inject}
\end{equation}

\subsubsection{Intra-partition self-attention}

For each frame $t$, joints in partition $r$ are treated as tokens, yielding $\mathbf{F}^{r}(t)\in\mathbb{R}^{B\times V_r\times C}$. Query, key, and value projections are computed as

\begin{equation}
\begin{aligned}
\mathbf{Q}^{r}(t)&=\mathbf{F}^{r}(t)\mathbf{W}_{Q}^{r},\\
\mathbf{K}^{r}(t)&=\mathbf{F}^{r}(t)\mathbf{W}_{K}^{r},\\
\mathbf{V}^{r}(t)&=\mathbf{F}^{r}(t)\mathbf{W}_{V}^{r},
\end{aligned}
\end{equation}

followed by

\begin{equation}
\widetilde{\mathbf{F}}^{r}(t)
=
\mathrm{Softmax}\left(
\frac{\mathbf{Q}^{r}(t)\mathbf{K}^{r}(t)^{\mathsf{T}}}{\sqrt{d}}
\right)\mathbf{V}^{r}(t).
\label{eq:ipsa}
\end{equation}

In the attention computation, $d$ denotes the feature (projection) dimension. Softmax is applied over joint tokens within the same partition. Refined features are written back to their original joint positions. This operation constrains attention to anatomically related joints at each frame and does not introduce temporal self-attention.

\subsubsection{Inter-partition fusion and classification}

Refined partition features are concatenated along the channel dimension and projected back to the original feature dimension:

\begin{equation}
\mathbf{X}^{\star}
=
\mathcal{P}\left(
\mathrm{Concat}_{C}
\left[
\widetilde{\mathbf{X}}^{u},
\widetilde{\mathbf{X}}^{m},
\widetilde{\mathbf{X}}^{l}
\right]
\right),
\label{eq:ipf}
\end{equation}

where $\mathcal{P}$ consists of a linear projection, Gaussian error linear unit activation, and layer normalization. The fused sequence is passed to the CTR-GCN backbone, global average pooling, and a fully connected layer that outputs probabilities for the three study-defined classes.

\subsection{Training, cross-validation, and baseline comparison}

Models were implemented in PyTorch and trained on a Linux workstation with an NVIDIA RTX 4090 GPU. Adam optimization used an initial learning rate of $1\times10^{-4}$, batch size of eight, cross-entropy loss, and a maximum of 100 epochs. ST-GCN \cite{Yan2018STGCN}, 2s-AGCN \cite{Shi2019AGCN}, MS-G3D \cite{Liu2020MSG3D}, HD-GCN \cite{Lee2023HDGCN}, InfoGCN \cite{Chi2022InfoGCN}, and CTR-GCN \cite{Chen2021CTRGCN} were evaluated as baselines.

The model was trained using participant-independent five-fold cross-validation. Input sequences were standardized to 50 frames with 32 skeletal joints and three coordinate channels (x, y, z). A fixed random seed (42) was used to ensure reproducibility. Early stopping was applied with a patience of 10 epochs, and the model checkpoint achieving the highest evaluation accuracy within each fold was retained.

\subsection{Evaluation and statistical analysis}

The primary outcome was three-class participant-level classification. Accuracy and macro-averaged precision, recall, and F1 score were calculated from pooled out-of-fold predictions. For class $i$, let $TP_i$, $FP_i$, and $FN_i$ denote true positives, false positives, and false negatives. With $K=3$ classes and $N$ participants,

\begin{equation}
\mathrm{Accuracy}=\frac{\sum_{i=1}^{K}TP_i}{N},
\end{equation}

\begin{equation}
\mathrm{Macro\mbox{-}Precision}
=
\frac{1}{K}\sum_{i=1}^{K}\frac{TP_i}{TP_i+FP_i},
\end{equation}

\begin{equation}
\mathrm{Macro\mbox{-}Recall}
=
\frac{1}{K}\sum_{i=1}^{K}\frac{TP_i}{TP_i+FN_i},
\end{equation}

\begin{equation}
\mathrm{Macro\mbox{-}F1}
=
\frac{1}{K}\sum_{i=1}^{K}
\frac{2TP_i}{2TP_i+FP_i+FN_i}.
\end{equation}

To examine the clinical relevance of the skeletal gait data, interpretable kinematic features were extracted at the participant level, and associations with continuous HAMD-17 total scores were evaluated using Spearman's rank correlation. Nominal $P$-values were adjusted using the Benjamini--Hochberg false discovery rate (BH-FDR) procedure. Pearson correlation coefficients and bootstrap 95\% confidence intervals were also calculated for selected features. Exploratory $\eta^2$ values were used to assess between-group kinematic separation, and within-group sensitivity analyses were performed within each severity group.

\section{Data availability}

The participant-level skeletal gait recordings and associated clinical data supporting the findings of this study are not publicly available because they contain sensitive behavioral and clinical information from adolescent participants. Access to these data is restricted by ethical and institutional requirements.

\section{Code availability}

The custom code used for skeletal gait preprocessing, UMLPA-GCN model training, and evaluation is not publicly available but may be made available to qualified researchers upon reasonable request to the corresponding authors, subject to institutional approval.

\section{Acknowledgements}

This work was supported in part by the National Natural Science Foundation of China (Grant No. 62573271), the Major Basic Research Project of the Natural Science Foundation of Shandong Province (Grant No. ZR2025ZD24), and the Chinese University of Hong Kong Vice-Chancellor Early Career Professorship Scheme (Grant No. 441).

\section{Author contributions}

Shuai Liu: Writing -- original draft, Writing -- review and editing, Conceptualization, Data curation. Xiang Li: Writing -- original draft, Writing -- review and editing. Guangxin Dai: Writing -- original draft, Writing -- review and editing, Formal analysis, Visualization. Mengyuan Yu: Writing -- original draft, Writing -- review and editing, Investigation, Data curation. Wenkong Wang: Writing -- original draft, Writing -- review and editing. Hanchi Liu: Writing -- original draft, Writing -- review and editing. Jiahui Niu: Writing -- original draft, Writing -- review and editing. Yujie Nie: Writing -- original draft, Writing -- review and editing. Yumei Wang: Writing -- original draft, Writing -- review and editing, Supervision, Data curation. Xin Ma: Writing -- original draft, Writing -- review and editing, Supervision, Funding acquisition. Lizhou Fan: Writing -- original draft, Writing -- review and editing, Conceptualization, Supervision, Funding acquisition.

\section{Competing interests}

The authors declare no competing interests.

\section{Ethics declarations}

The study was approved by the Ethics Committee of Shandong Provincial Hospital Affiliated to Shandong First Medical University (approval no. R202411120385) and was conducted in accordance with the Declaration of Helsinki. Written informed consent was obtained from participants and their legal guardians before study procedures, as described in the Methods.

\clearpage
\appendix
\section*{Appendix A: Methods}

\subsection*{Method S1. Study population and eligibility}

Participants were recruited from the Clinical Psychology Ward of Shandong Provincial Hospital Affiliated to Shandong First Medical University. For the depression groups, participants were required to meet the DSM-5 diagnostic criteria for major depressive disorder and to be currently experiencing a depressive episode. They also had to have received no treatment (pharmacological or psychotherapeutic) within the preceding 14 days. In addition, they had to have completed at least primary school education, with sufficient ability to understand the study procedures and the rating scales. Exclusion criteria included a current or past major psychiatric disorder other than depression, a serious organic disease affecting any major system or organ, participation in an interventional clinical trial within the preceding three months, poor compliance, or any other condition judged by the investigators to make the participant unsuitable for the study. Healthy controls were recruited from a comparable age range and screened to exclude major psychiatric disorders and severe organic diseases.

Depression severity was assessed with the 17-item Hamilton Depression Rating Scale (HAMD-17), administered by trained psychiatrists. Participants with HAMD-17 scores of 0--7 were categorized as healthy controls (HC), those with scores of 8--16 as mild depression (MD), and those with scores of 17 or higher as moderate-to-severe depression (MSD). These bands follow the empirically derived severity cutoffs recommended for the 17-item scale (no depression 0--7, mild 8--16, moderate 17--23, severe $\geq$24; M.~Zimmerman, J.~H. Martinez, D.~Young, I.~Chelminski, and K.~Dalrymple, ``Severity classification on the Hamilton Depression Rating Scale,'' \emph{Journal of Affective Disorders}, vol.~150, no.~3, pp.~384--388, 2013), rather than the more commonly used clinical split (8--13/14--18). Although scores of 17 or higher are commonly further split into moderate and severe categories in clinical practice, these two categories were merged in this study because of sample imbalance; this grouping produced the three-level severity classification used throughout the analysis.

\subsection*{Method S2. Gait data acquisition and preprocessing}

Gait recordings were collected in a quiet hospital corridor within the Clinical Psychology Ward, under relatively controlled conditions. Ceiling-mounted Azure Kinect depth sensors captured 32-joint skeleton sequences. Each participant completed two walking trials along a predefined route at a self-selected habitual walking speed; two clinicians stood at opposite ends of the corridor to indicate the turning points, monitor the participant, and assist if needed. Participants were asked to walk as naturally as possible, with short breaks between trials to limit fatigue.

Raw skeleton sequences underwent four preprocessing steps: (i) camera-angle correction, (ii) root-joint normalization, (iii) Gaussian smoothing, and (iv) temporal standardization to 50 frames. Step (i) estimated the tilt angle between the camera optical axis and the ground plane from the SpineBase trajectory. Step (ii) subtracted the SpineBase coordinates from all joints in every frame. Step (iii) applied a Gaussian smoothing filter with a five-frame window and a standard deviation of 1.0 frame to each coordinate trajectory independently. Step (iv) retained the first 50 frames of sequences longer than 50 frames, discarding the subsequent frames, zero-padded sequences shorter than 50 frames to reach 50 frames. A target length of 50 frames was chosen based on the observed distribution of retained gait-segment lengths, most of which were approximately 40--60 frames.

\subsection*{Method S3. Cross-validation and training protocol}

All models were implemented in PyTorch and trained on a single NVIDIA RTX 4090 GPU. Participant-independent five-fold cross-validation was used. First, the 119 participants were grouped into healthy controls (HC, $n=39$), mild depression (MD, $n=41$), and moderate-to-severe depression (MSD, $n=39$). Participants in each severity group were divided into five subsets; subsets with the same fold index from the three groups were then combined to form each of the five folds. In each iteration, one combined fold served as the held-out test set and the remaining four folds served as the training set. The fold assignments were fixed across model evaluations. Each participant was assigned to exactly one test fold, and all recordings, cameras, walking trials, and derived segments belonging to that participant remained in the same fold. Primary metrics were computed from pooled out-of-fold participant predictions, such that each participant contributed one final prediction.

Optimization used Adam with an initial learning rate of $1\times10^{-4}$, a batch size of eight, and cross-entropy loss, for a maximum of 100 epochs. A fixed random seed of 42 was used for model training. The UMLPA-GCN backbone comprised $K=10$ stacked CTR-GCN blocks, followed by global average pooling, a fully connected layer, and a softmax classifier over the three severity classes.

\subsection*{Method S4. Participant-level kinematic feature extraction and statistical analysis}

Interpretable kinematic features were extracted from the retained frontal-view skeletal sequences and aggregated at the participant level, yielding one value per participant for each available feature. Skeleton frames were sorted by their timestamps, and joint coordinates associated with duplicate timestamps were averaged. Missing values in coordinate trajectories were linearly interpolated according to frame index, with the nearest valid observations carried forward or backward at the two sequence boundaries. For the kinematic feature analysis, coordinate trajectories were subsequently smoothed using a three-frame median filter. Sequence duration was calculated as the difference between the first and last timestamps. The analysis included 119 participants: 39 healthy controls (HC), 41 with mild depression (MD), and 39 with moderate-to-severe depression (MSD).

For each frame, the midpoint between the left and right shoulders and the midpoint between the left and right hips were calculated. The median distance between these two midpoints across the sequence was used as the trunk-length normalization scale $L$. The hip midpoint was used as the origin of the relative coordinate system, with the $z$-axis serving as the reference vertical direction. A fixed left--right direction was constructed from the median orientation of the line connecting the shoulders across the sequence, and the corresponding forward direction was then defined. For a time-varying quantity $x$, movement amplitude and joint range of motion (ROM) were estimated using the difference between the 95th and 5th percentiles,
\begin{equation*}
R(x)=Q_{95}(x)-Q_{5}(x).
\end{equation*}

Fifteen predefined features were extracted: pelvis path speed, hip forward speed proxy, mean absolute trunk lean, trunk lean variability, head movement range, head movement root mean square (RMS), left arm swing range, right arm swing range, arm swing asymmetry, foot movement asymmetry, mean knee ROM, mean elbow ROM, mean hip ROM, cadence proxy, and step length proxy. Pelvis path speed was calculated as the sum of the three-dimensional displacements of the pelvis between consecutive frames divided by sequence duration and then normalized by $L$ (body lengths/s). The hip forward speed proxy was calculated as the sum of absolute frame-to-frame changes in the forward coordinate of the pelvis relative to the hip midpoint, divided by duration, and was retained in source coordinate units/s. Trunk lean was derived from the forward and reference-vertical components of the vector extending from the hip midpoint to the shoulder midpoint. Its mean absolute value and standard deviation were used as two separate features (degrees).

Head position was defined as the mean of the retained head and facial joint coordinates. Head movement range was calculated as the Euclidean norm of the three axis-specific ranges $R(x)$ of the head position relative to the hip midpoint, divided by $L$ (body lengths). Head movement RMS was calculated after subtracting the temporal mean from the relative head-position trajectory, taking the root mean square over all frames and all three coordinate components, and dividing by $L$ (body lengths). Left and right arm swing ranges were separately calculated as $R(x)$ of the wrist's forward position relative to the hip midpoint, divided by $L$; the corresponding foot movement amplitudes were calculated in the same way. Arm swing asymmetry and foot movement asymmetry were defined as
\begin{equation*}
\operatorname{Asymmetry}(\%)=
100\,\frac{|A_R-A_L|}{(A_R+A_L)/2},
\end{equation*}
where $A_R$ and $A_L$ denote the right- and left-side movement amplitudes, respectively. Knee and elbow angles were defined by the hip--knee--ankle and shoulder--elbow--wrist joint triplets, respectively. The hip-angle proxy was defined by the lower-spine--hip--knee joint triplet. For each of these joints, the side-specific ROM was estimated using $R(x)$ of the joint-angle trajectory (degrees), and the left- and right-side ROMs were averaged.

Cadence and step length proxies were derived from local extrema in the left and right feet's relative forward-motion trajectories. The minimum event separation was set to approximately 0.15\,s and at least two frames. The cadence proxy was calculated as the number of retained events divided by sequence duration (Hz). The step length proxy was the median absolute difference between the forward foot positions at successive contralateral events, normalized by $L$ (body lengths).

Associations between each of the 15 kinematic features and the continuous HAMD-17 total score were evaluated using two-sided Spearman rank correlation tests. Spearman coefficients $\rho$ and unadjusted $P$-values were reported. The 15 tests were jointly corrected using the Benjamini--Hochberg false discovery rate (BH-FDR) procedure, with $q<0.05$ considered statistically significant. Pearson correlation coefficients were also calculated to describe linear associations. Participant-level bootstrap 95\% confidence intervals were estimated by resampling paired feature and HAMD-17 observations with replacement 1,000 times, keeping the original sample size in each resample; the 2.5th and 97.5th percentiles of valid bootstrap estimates defined the confidence limits. Each test used participants with both the feature and the HAMD-17 score available. The step length proxy had 118 valid participants, whereas each of the other 14 features had 119.

Exploratory between-group effect sizes were quantified using eta squared ($\eta^2$):
\begin{equation*}
\eta^2=\frac{SS_{\mathrm{between}}}{SS_{\mathrm{total}}},\qquad
SS_{\mathrm{between}}=\sum_g n_g(\bar{x}_g-\bar{x})^2,\qquad
SS_{\mathrm{total}}=\sum_i(x_i-\bar{x})^2,
\end{equation*}
where $n_g$ denotes the number of valid participants in group $g$, and $\bar{x}_g$ and $\bar{x}$ are the corresponding group mean and overall mean. Within-group sensitivity analyses repeated Spearman correlation testing in the HC, MD, and MSD groups separately, with BH-FDR adjustment applied independently to the 15 feature tests within each group. In addition, exploratory linear regression models were fitted with HAMD-17 total score as the dependent variable and a standardized kinematic feature and group-label dummy variables as predictors. Feature regression coefficients and their 95\% confidence intervals were estimated, and the 15 feature-specific regression tests were adjusted using the BH-FDR procedure.

\clearpage

\section*{Appendix B: Tables}

\subsection*{Table S1. Detailed demographic and clinical characteristics}

Group-level statistics corresponding to the main-text Table 1 are shown below.

\begin{table}[h]
\centering
\small
\begin{tabular}{lccc}
\toprule
Characteristic & HC ($n=39$) & MD ($n=41$) & MSD ($n=39$) \\
\midrule
Sex, male/female & 15/24 & 13/28 & 12/27 \\
Age, years (mean$\pm$SD) & $17.11\pm2.16$ & $16.34\pm2.02$ & $16.54\pm2.09$ \\
HAMD-17 total (mean$\pm$SD) & $1.95\pm2.29$ & $12.00\pm1.73$ & $23.38\pm2.54$ \\
\bottomrule
\end{tabular}
\end{table}

\clearpage

\subsection*{Table S2. Azure Kinect 32-joint mapping and partition membership}

The Azure Kinect body-tracking model provides 32 joints (indices 0--31). The table below lists each joint and the anatomical partition to which it is assigned in UMLPA-GCN; the partition masks $\mathbf{M}^{u},\mathbf{M}^{m},\mathbf{M}^{l}$ are derived from this assignment.

\begin{table}[H]
\centering
\small
\setstretch{1.0}
\renewcommand{\arraystretch}{0.9}
\begin{tabular}{clll}
\toprule
Index & Joint & Parent & Partition \\
\midrule
0  & SPINE\_BASE     & ---               & Lower \\
1  & SPINE\_NAVAL    & SPINE\_BASE       & Middle \\
2  & SPINE\_CHEST    & SPINE\_NAVAL      & Middle \\
3  & NECK            & SPINE\_CHEST      & Upper \\
4  & CLAVICLE\_L     & SPINE\_CHEST      & Middle \\
5  & SHOULDER\_L     & CLAVICLE\_L       & Middle \\
6  & ELBOW\_L        & SHOULDER\_L       & Middle \\
7  & WRIST\_L        & ELBOW\_L          & Middle \\
8  & HAND\_L         & WRIST\_L          & Middle \\
9  & HANDTIP\_L      & HAND\_L           & Middle \\
10 & THUMB\_L        & WRIST\_L          & Middle \\
11 & CLAVICLE\_R     & SPINE\_CHEST      & Middle \\
12 & SHOULDER\_R     & CLAVICLE\_R       & Middle \\
13 & ELBOW\_R        & SHOULDER\_R       & Middle \\
14 & WRIST\_R        & ELBOW\_R          & Middle \\
15 & HAND\_R         & WRIST\_R          & Middle \\
16 & HANDTIP\_R      & HAND\_R           & Middle \\
17 & THUMB\_R        & WRIST\_R          & Middle \\
18 & HIP\_L          & SPINE\_BASE       & Lower \\
19 & KNEE\_L         & HIP\_L            & Lower \\
20 & ANKLE\_L        & KNEE\_L           & Lower \\
21 & FOOT\_L         & ANKLE\_L          & Lower \\
22 & HIP\_R          & SPINE\_BASE       & Lower \\
23 & KNEE\_R         & HIP\_R            & Lower \\
24 & ANKLE\_R        & KNEE\_R           & Lower \\
25 & FOOT\_R         & ANKLE\_R          & Lower \\
26 & HEAD            & NECK              & Upper \\
27 & NOSE            & HEAD              & Upper \\
28 & EYE\_L          & HEAD              & Upper \\
29 & EAR\_L          & HEAD              & Upper \\
30 & EYE\_R          & HEAD              & Upper \\
31 & EAR\_R          & HEAD              & Upper \\
\bottomrule
\end{tabular}
% \caption*{\footnotesize Partition membership as implemented in UMLPA-GCN (confirmed against the analysis pipeline).}
\end{table}

\clearpage

\subsection*{Table S3. Training hyperparameters and implementation}

\begin{table}[H]
\centering
\small
\begin{tabular}{ll}
\toprule
Item & Value \\
\midrule
Framework & PyTorch \\
GPU & NVIDIA RTX 4090 (single) \\
Optimizer & Adam \\
Initial learning rate & $1\times10^{-4}$ \\
Batch size & 8 \\
Loss & Cross-entropy \\
Maximum epochs & 100 \\
CTR-GCN blocks ($K$) & 10 \\
Input & 32 joints $\times$ 3 coordinate channels $\times$ 50 frames \\
Readout & global average pooling + fully connected + softmax (3 classes) \\
Learning-rate schedule & Constant (no scheduler) \\
Weight decay & 0 \\
Random seed & 42 \\
\bottomrule
\end{tabular}
% \caption*{\footnotesize Hyperparameters for the primary analysis.}
\end{table}

\clearpage

\subsection*{Table S4. Model architecture and implementation details}

\begin{table}[H]
\centering
\small
\begin{tabular}{ll}
\toprule
Component & Description \\
\midrule
UMLPM & divides the 32-joint skeleton into upper, middle, and lower partitions (Table S2) \\
PPE & Partition Positional Encoding: partition-aware positional information \\
IPSA & Intra-Partition Self-Attention: adaptive joint refinement within each partition \\
IPF & Inter-Partition Fusion: channel-wise concatenation + linear + GELU + layer norm \\
Backbone & 10 CTR-GCN blocks \\
Readout & global average pooling + fully connected + softmax (3) \\
\bottomrule
\end{tabular}
\end{table}

\clearpage

\section*{Appendix C: Figures}

\renewcommand{\figurename}{Supplementary Fig.}
\setcounter{figure}{0}
\renewcommand{\thefigure}{S\arabic{figure}}

\begin{figure}[h]
\centering
\includegraphics[width=\textwidth]{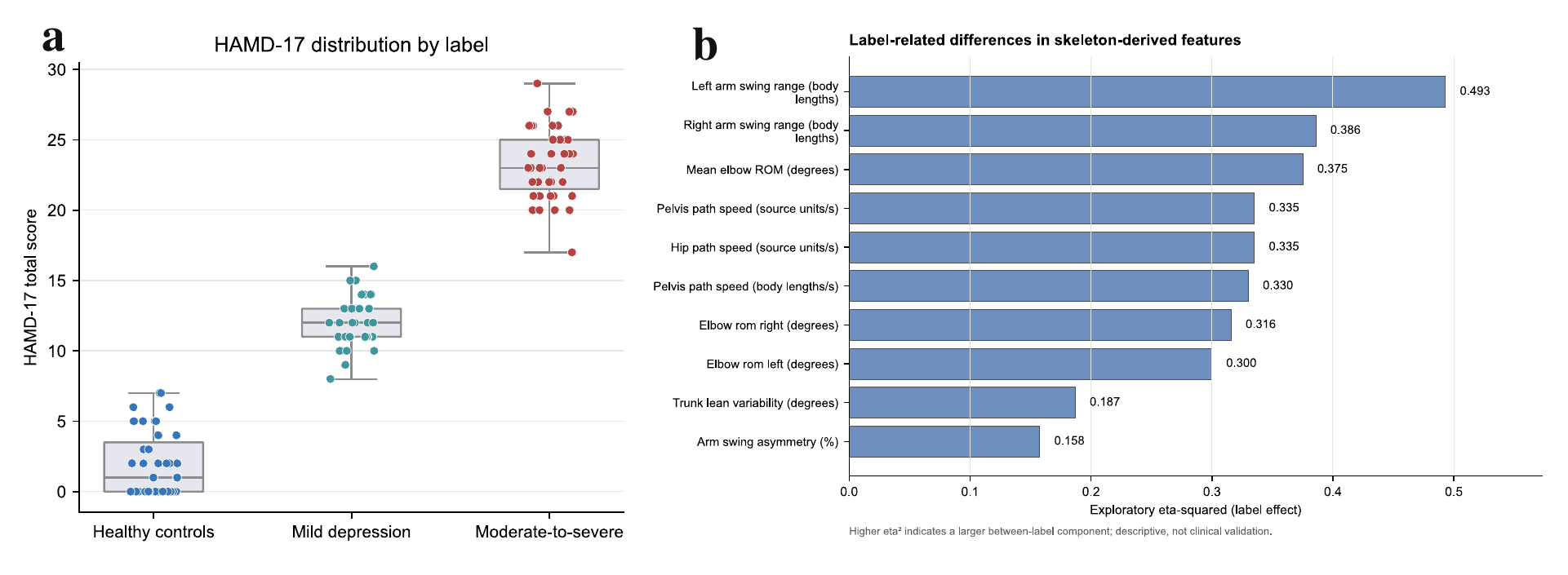}
\caption{Clinical severity stratification and exploratory between-group kinematic effect sizes. (a) Distribution of HAMD-17 total scores across healthy controls (HC), mild depression (MD), and moderate-to-severe depression (MSD). The observed separation is consistent with the predefined HAMD-17-based severity thresholds. (b) Exploratory $\eta^2$ ranking of the predefined kinematic features across the three severity groups; higher $\eta^2$ indicates a larger proportion of variance associated with between-group differences.}
\end{figure}

\begin{figure}[h]
\centering
\includegraphics[width=\textwidth]{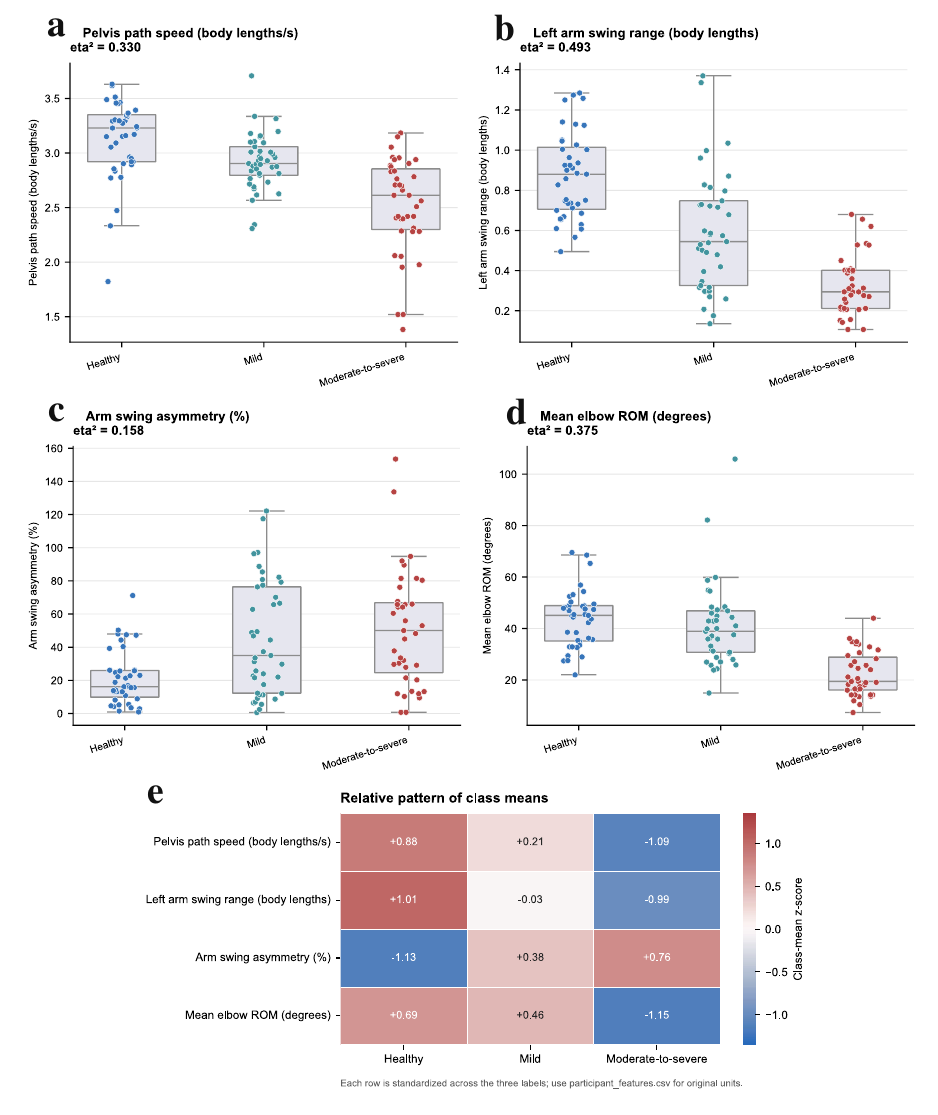}
\caption{Group-wise distributions of representative kinematic features across HC, MD, and MSD. (a) Pelvis path speed. (b) Left arm swing range. (c) Arm swing asymmetry. (d) Mean elbow range of motion (ROM). (e) Standardized group-mean profiles of the four representative features. Each point in (a)--(d) represents one participant.}
\end{figure}

\begin{figure}[h]
\centering
\includegraphics[width=\textwidth]{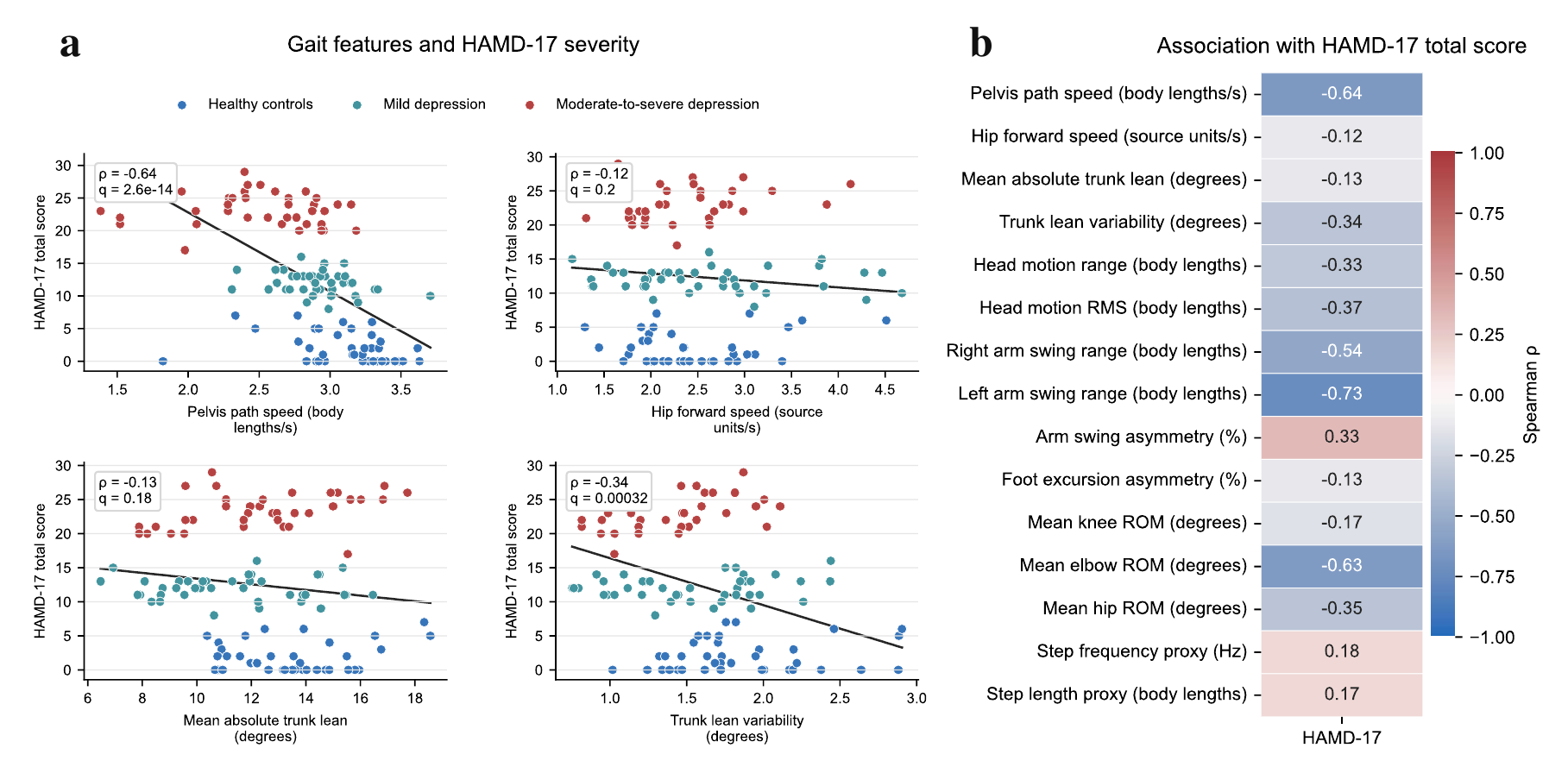}
\caption{Participant-level associations between representative kinematic features and HAMD-17 total score. (a) Scatter plots for left arm swing range, pelvis path speed, mean elbow ROM, and arm swing asymmetry; each point is one participant ($n=119$) and solid lines are fitted linear trends for visualization only (inference was based on Spearman's rank correlation). (b) Spearman correlation coefficients between HAMD-17 total score and the 15 predefined kinematic features; asterisks mark features that remained significant after BH-FDR correction.}
\end{figure}

\end{document}